\documentclass[useAMS,usenatbib]{mn2e}
\usepackage{graphicx}
\usepackage{amsmath}
\usepackage{amssymb}

\usepackage{dcolumn}
\usepackage{bm}

\title[Interaction of clumpy dark matter with gas]{Interaction
of clumpy dark matter with interstellar medium in astrophysical systems.}

\author[A. N. Baushev]{A. N. Baushev\\
Bogoliubov Laboratory of Theoretical Physics, Joint Institute for Nuclear Research,
141980 Dubna, Moscow Region, Russia\\
DESY, 15738 Zeuthen, Germany\\
 Institut f\"ur Physik und Astronomie, Universit\"at Potsdam, 14476
Potsdam-Golm, Germany\\}
\begin{document}

\date{}

\pagerange{\pageref{firstpage}--\pageref{lastpage}} \pubyear{2011}

\label{firstpage}

\maketitle

\begin{abstract}
Contemporary cosmological conceptions suggest that the dark matter in haloes of galaxies and galaxy
clusters has most likely a clumpy structure. If a stream of gas penetrates through it, a
small-scale gravitational field created by the clumps disturbs the flow resulting in momentum
exchange between the stream and the dark matter. In this article, we perform an analysis of this
effect, based on the hierarchial halo model of the dark matter structure and Navarro-Frenk-White
density profiles. We consider the clumps of various masses, from the smallest up to the highest
ones $M\ge 10^{9} M_\odot$. It has been found that in any event the effect grows with the mass of
the clump: not only the drag force $\mathfrak F$ acting on the clump, but also its acceleration
$w=\mathfrak F/M$ increases.

We discuss various astrophysical systems. The mechanism proved to be ineffective in the case of
galaxy or galaxy cluster collisions. On the other hand, it played an important role during the
process of galaxy formation. As a result, the dark matter should have formed a more compact,
oblate, and faster rotating substructure in the halo of our Galaxy. We have shown that this thick
disk should be more clumpy than the halo. This fact is very important for the indirect detection
experiments since it is the clumps that give the main contribution to the annihilation signal. Our
calculations show that the mechanism of momentum exchange between the dark and baryon matter is
ineffective on the outskirts of the galactic halo. It means that the clumps from there were not
transported to the thick disk, and this regions should be more clumpy than the halo on the average.
\end{abstract}

\begin{keywords}
cosmology: dark matter, cosmology: theory, elementary particles.
\end{keywords}

\section{Introduction}
In modern representation about 80\% of the matter forming structures in the Universe is dark matter
(DM). In particular, it is the dark matter that makes the main contribution to the contents of
galactic haloes. The nature of the dark matter is presently unknown. It is widely believed that it
consists of some weakly or extremely weakly interacting particles generated in the early Universe.
Modern cosmological observations \citep{wmap} disclose that the dark matter is cold (CDM).

Direct detection experiments \citep{bertone2005} impose strict upper constraints on the DM
particle-nucleon scattering cross section. Therefore, the dark matter presence does not affect
normal matter propagation through it; it only makes a contribution to the total mass of the Galaxy
and consequently to the large-scale gravitational field.

However, this conclusion is valid only if the dark matter distribution is uniform. At the same
time, contemporary cosmological conceptions suggest that the dark matter has very likely a clumpy
structure.  If the dark matter contains clumps, their small-scale gravitational field exerts extra
influence on normal substance. For instance, if a stream of gas flows through clumpy dark matter,
the field disturbs the flow. It results in momentum exchange between the stream and the dark matter
and partial transformation of the kinetic energy of the stream into heat.

The dark matter perturbations played an important role in the universe structure formation
\citep{gorbrub2}. The clumps were formed in the early Universe from some primordial fluctuations,
and their present-day mass distribution depends on the fluctuation spectrum that is not known very
well. Inflation theory predicts that the spectrum had a flat Harrison-Zeldovich shape
\citep{gorbrub2}; so perturbations with all masses existed in the early Universe. However, the
clump mass distribution may have a cut-off in the area of small masses (very small clumps should be
destroyed by free-streaming). The minimal possible clump mass depends on the physical nature of the
dark matter particles (especially on the mass). For one of the most popular dark matter candidates
--- neutralino, the lightest SUSY particle
--- the limit estimations vary from $10^{-12} M_\odot$ \citep{10-12} to $10^{-6} M_\odot$ \citep{10-6}. However, clumps
with masses near the limit should prevail, for the number of clumps grows with reduction in the mass.

Another important process strongly affecting the dark matter structure is the tidal destruction of
small clumps by the bigger ones \citep{berezinsky2006}. As a consequence, a significant part of
dark matter supposedly is not included in clumps forming a more or less uniform component. The
presence of the uniform component in the dark matter distribution does not influence, of course,
the small-scale structure of the gravitational field that is totaly created only by density
perturbations over the minimal level (i.e. by clumps). On the other hand, it is
 shown \citep{berezinsky2006} that the densest central part of small clumps can hardly be destroyed
by the tidal forces.

Hence if the dark matter is really composed of WIMPs, the presence of clumps of various masses down
to very small is a necessity rather than a theoretical possibility. The  biggest clumps can be
directly observed: these are nothing else than galactic haloes. Smaller clumps are presently beyond
experimental detection. Meanwhile, any evidence for their existence could be very important: for
instance, if the low-mass clumps are found to be lacking, it will probably mean that the dark
matter is not composed of WIMPs. The second reason while the presence of small clumps is important
is that if the dark matter annihilation is possible, it goes on mainly in clumps
\citep{berezinsky2006}. In realistic cosmological models low-mass clumps collapse earlier and turn
out to be significantly denser than the massive ones, and it is these clumps that give the main
contribution to the annihilation signal. Thus, the space distribution of low-mass clumps is very
important for indirect detection experiments.

As we will see, the drag effect acting on the clump always grows with the clump mass, and the
distribution of the smallest clumps can hardly be disturbed by any real astrophysical gas stream.
On the other hand, clumps form a sort of hierarchical structure, in which small objects are bound
in the gravitational field of the large ones. As a result, small clumps can be dragged together
with the heavy objects to which they belong.

\section{The structure of clumpy dark matter}
First of all, density and gravitational potential distributions of the clumpy matter need to be
ascertained. Unfortunately, the problem remains to be solved. Numerous models of the dark matter
structure and clump density profiles are considered in literature. In this article, we will use the
results of \citep{bullock} as one of possible scenarios, escaping the discussion of which of the
models of the structure formation is the most plausible, which is far beyond the scope of this
work. Moreover, as we will see, the result is not very sensitive to the parameters of the density
profile.

In accordance with \citep{bullock}, we adopt the Navarro-Frenk-White density profile of a clump
\begin{equation}
\label{a1}
 \rho_{\scriptscriptstyle NFW}= \dfrac{\rho_s}{(r/r_s)(1+r/r_s)^2}
\end{equation}
where $\rho_s$ and $r_s$ are characteristic 'inner' density and radius of the clump. It is worth mentioning that
profile (\ref{a1}) cannot be valid for all $r$, as the total clump mass diverges when $r\to\infty$.

Instead of $\rho_s$ and $r_s$ it is convenient to use the virial radius $R_{vir}$, mass $M_{vir}$, and the
concentration parameter
 \begin{equation}
 c_{vir}\equiv \dfrac{R_{vir}}{r_s}
 \label{a2}
 \end{equation}
The above-mentioned quantities are related by the equations (see \citep{bullock} for details)
 \begin{align}
 M_{vir} &=\dfrac{4\pi}{3}\Delta_{vir} \rho_u R^3_{vir} \label{a3}\\
 M_{vir} &=4\pi \rho_s r^3_s A(c_{vir}) \label{a4}\\
 A(c_{vir}) &\equiv \ln(c_{vir}+1)-\dfrac{c_{vir}}{c_{vir}+1}\nonumber
 \end{align}
Here $\rho_u$ is the mean universe density, $\Delta_{vir}$ is the virial overdensity. For the present epoch
$\Delta_{vir}\simeq 337$. From equations (\ref{a2}-\ref{a4}) we can see that among the quantities $\rho_s$, $r_s$
$R_{vir}$, $M_{vir}$, and $c_{vir}$ there are only two independent. It is convenient to use $M_{vir}$ and $c_{vir}$.

The clump potential is:
 \begin{equation}
 \label{c1}
 \phi(r) = -\dfrac{G M_{vir}}{r_s A(c_{vir})} \dfrac{r_s}{r}\ln(\dfrac{r}{r_s}+1)
 \end{equation}
It is worth noting that, contrary to the Newtonian case, the potential remains finite at the centre of the clump.
 \begin{equation}
 \label{c2}
 \phi_0 = -\dfrac{G M_{vir}}{r_s A(c_{vir})}
 \end{equation}

Equation (\ref{c1}) for the clump gravitational field cannot be valid for an arbitrary big radius
$r$. First, as we have already mentioned, the total mass of a clump with profile (\ref{a1})
diverges when $r\to\infty$. Second, equation (\ref{c1}) is certainly inapplicable at a radius $r$
if $ M < \frac{4\pi}{3}\bar \rho r^3$, where $M$ is the clump mass and $\bar \rho$ is the dark
matter density averaged over a region vastly larger than the clump size (for instance, on the
outskirts of the Solar System $\bar \rho \simeq 0.3\, \mbox{GeV}\!/\mbox{cm}^3$ \citep{gorbrub1}).
Indeed, this condition means that the matter not included into the clump already prevails inside
the sphere of radius $r$ circling the centre of the clump and gives a bigger contribution to the
gravitational field at this radius. From the latter consideration, we can roughly estimate the
radius $R_\Xi$ of the "sphere of influence" of a clump, where its gravitational field of the clump
dominates, and equation (\ref{c1}) is applicable $ R_\Xi = \root \displaystyle{3} \of
{\frac{3M}{4\pi\bar \rho}}$. By the clump mass we hereafter imply its virial mass $M_{vir}$. It may
appear strange that we use $M_{vir}$ while $R_\Xi$ can be significantly smaller than $R_{vir}$.
However, it is easy to see from (\ref{a1}) that the $r$-dependence of the mass is only
logarithmical, and $M(R_\Xi)$ does not significantly differ from $M_{vir}$ if $R_\Xi\gg r_s$.
Moreover, profile (\ref{a1}) obviously does not hold for $r>R_\Xi$ because of the tidal
perturbations, and the question of exact density distribution is extremely complex. Consequently,
supposition $M\equiv M_{vir}$ is quite acceptable in our approximative calculations. It is
convenient to use
 \begin{equation}
 \label{a11}
 \Xi\equiv \dfrac{R_\Xi}{r_s} = \root \displaystyle{3} \of {\dfrac{3M_{vir}}{4\pi\bar \rho r^3_s}} =
 c_{vir} \root \displaystyle{3} \of {\dfrac{\Delta_{vir} \rho_u}{\bar \rho}}
 \end{equation}
instead of $R_\Xi$.

\section{Clump interaction with a stream of gas}

The nature of the clump interaction with a stream of gas depends on the relation between $R_\Xi$
and the free length $l_{fl}$ of the gas atoms. We start our consideration from the case when $R_\Xi
\ll l_{fl}$. Then we can look upon the gas flow as being a stream of noninteracting particles and
consider propagation of each atom separately. A dynamical friction between the clump and the stream
appears as a result of the particle scattering in the gravitational field of the clump \citep{chan,
binneytremaine}.

Let us consider a stream of particles of mass $m$ moving at a speed $\upsilon_\infty$ at infinity
and scattered on potential (\ref{c1}). The total momentum change of the gas stream in a unit time
(i.e., the force $\mathfrak F$ acting on the clump) is:
 \begin{equation}
 \label{a18}
  \mathfrak F =\int^{R_\Xi}_{0}\!\! \Delta p_x\cdot n\upsilon_\infty\cdot 2\pi \varpi d\varpi
 \end{equation}
Here $\varpi$ is the impact parameter of the trajectory and is $n$ is the gas concentration. The
problem can be simplified if we take into account that the clump gravitational field is
respectively weak, and the scattering angle $\theta$ is small. Then after simple calculations we
obtain for the clump acceleration $w\equiv \mathfrak F/M_{vir}$:
 \begin{equation}
 \label{b1}
 w = \dfrac{\pi \nu G^2 M_{vir}}{3\upsilon^2_\infty A^2(c_{vir})}\cdot\ln^3\left(\dfrac{c_{vir}}{2}\root \displaystyle{3} \of {\dfrac{\Delta_{vir} \rho_u}{\bar \rho}}\right)
 \end{equation}
 We introduced the gas density $\nu= m n$. One can see that the result is not very sensitive to the precise determination of $c_{vir}$ or
$\Xi$ while $c_{vir}\gg 1$ and $\Xi\gg 1$.

 In order to complete the calculations, we should define the dependence $c_{vir}(M_{vir})$ in an explicit form. Unfortunately, there are no
reliable estimations of this relationship in the range of small clump masses. \citep{bullock}
reported about power-law growth of halo concentration $c_{vir}$ with $M_{vir}$ decreasing, and the
results of \citep{ahn2005} confirm this conclusion. However, if we try to interpolate this
dependence to the area of the lowest possible masses, we obtain grotesquely huge values of
$c_{vir}$. It is likely that in this realm the relationship is somehow modified; however, halo
concentration should be very big ($c_{vir}\gg 1$). Following \citep{ahn2005}, we adopt that
$c_{vir}$ is constant $c_{vir}=70$ at low masses $M_{vir}\le 10^{10} M_{\odot}$ (though this
premise looks unlikely). Considering the uncertainty of $c_{vir}$ determination and the
above-mentioned weak dependence of the result on $c_{vir}$ and $\Xi$ we can neglect the difference
between $R_\Xi$ and $R_{vir}$ (i.e., we adopt $\bar \rho = \Delta_{vir} \rho_u$, which is quite
natural for galactic haloes). Then $c_{vir}=\Xi$, and it is easy to see that, if $c_{vir}\gg 1$,
$\frac{\ln^3(\Xi/2)}{A^2(c_{vir})}\simeq \ln\left(\dfrac{c_{vir}}{2}\right)$, and we can simplify
equation~(\ref{b1}):
 \begin{equation}
 \label{b3}
 w = \dfrac{\pi \nu G^2 M_{vir}}{3\upsilon^2_\infty}\cdot\ln\left(\dfrac{c_{vir}}{2}\right)
 \end{equation}
Formally, equation (\ref{b3}) rapidly grows when $\upsilon_\infty\to 0$. However, if
$\upsilon_\infty$ is small the scattering angle $\theta$ becomes large, and (\ref{b3}) is no longer
applicable. In order to estimate the maximum possible clump acceleration $w_{max}$, we should
perform a more detailed treatment.

In a general way, equation (\ref{a18}) can be rewritten as
\begin{equation}
 \label{a22}
  \mathfrak F = 2\pi \nu \upsilon^2_\infty \int^{R_\Xi}_{0}\!\! (1-\cos\theta) \varpi d\varpi
\end{equation}
Since $(1-\cos\theta)\le 2$ we can affirm that
\begin{equation}
 \label{a23}\mathfrak F\le 2\pi \nu \upsilon^2_\infty
R^2_\Xi
\end{equation}
On the other hand, when $\upsilon_\infty$
  is very small, the scattering
angle is large, and $(1-\cos\theta)$ is not a small number. Therefore, equation (\ref{a23}) can be
considered as an estimation of  $\mathfrak F$ in the limit of small velocity. Consequently, the
force grows as $\upsilon^2_\infty$ while $\upsilon_\infty$ is small and decreases as
$\upsilon^{-2}_\infty$ for big $\upsilon_\infty$. We can estimate $\upsilon^2_\infty$ that provides
the maximum of  $\mathfrak F$ equating expressions (\ref{b1}) and (\ref{a23}):
\begin{equation}
\label{a26}
 \upsilon^2_\infty  =
 \dfrac{G M_{vir}}{R_{vir}}\sqrt{\dfrac{\ln (c_{vir}/2)}{6}}
\end{equation}
Substituting this equation into (\ref{a23}), we obtain
\begin{equation}
\label{a27} w_{max}= 2 \pi\nu G R_{vir}\sqrt{\dfrac{\ln (c_{vir}/2)}{6}}
\end{equation}
Maximum clump acceleration weakly (as $\sqrt[3]{M_{vir}}$) depends on the clump mass, and if
$c_{vir}=70$ $w_{max}\simeq 4.8 \nu G  R_{vir}$.

In the case when $R_\Xi \ge l_{fl}$ the hydrodynamic approach should be used. The streamline
picture about the clump depends on the Mach number at infinity $M_M \equiv \upsilon_\infty / a$,
where $a$ is the sound speed. It is extremely difficult to calculate the flow in potential field
(\ref{c1}) exactly. However, the problem is thoroughly studied numerically, and we can easily
obtain simple estimations \citep{ostriker, brandenburg}. Beside this, we can use a similarity
between the system under consideration and gas accretion on a compact astrophysical object: the
later problem has been thoroughly studied \citep{lipunov}. The case $\upsilon_\infty < a$
corresponds to the spherical accretion \citep{Bondi52}, the case $\upsilon_\infty > a$ --- to the
cylindrical one \citep{BondiHoyle}. We emphasize, however, the above-mentioned important
distinction from the accretion problem: potential (\ref{c1}) is everywhere finite $|\phi|\le
|\phi_0|$.

If $\upsilon_\infty \gg a$ we may neglect the pressure. Then the field of flow velocities coincides
with the collisionless case, except for a narrow zone behind the clump, where the streamlines cross
and a shock wave appears (see \citep{BondiHoyle} for details). The form of the shock is not quite
clear, and we adopt the simplest supposition that it is a cone with the vertex at the centre of the
clump and the corner angle $\sim \frac{1}{M_M}=\frac{a}{\upsilon_\infty}$ \citep{lipunov}, which
corresponds to the solid angle $d\Omega=\pi \left(\frac{a}{\upsilon_\infty}\right)^2$. So all the
space around the clump can be divided on two regions: region $I$ before the shock wave, where the
stream velocity coincides with the collisionless case, and conic region $II$ after the shock wave.
Consequently, the total force acting on the clump (and consequently, its acceleration) is a sum of
its interactions with the flow in regions $I$ and $II$ ($w_1$ and $w_2$, respectively). In order to
calculate $w_1$ we may use equation (\ref{b3}) as in the collisionless case.

To estimate $w_2$ we assume that the shock wave is strong. Then the gas after the shock wave is
$\frac{\gamma+1}{\gamma-1}$ times denser than before, and the overdensity in region $II$ is
$\nu_{over}=\frac{2}{\gamma-1} \nu$. The total force of attraction $\mathfrak F_2$ between the
clump and the substance behind the wave can be roughly estimated as
\begin{equation}
 \label{d2}
\mathfrak F_2 = \int^{R_\Xi}_{0}\!\! G\dfrac{M_{vir} dm}{r^2}
\end{equation}
where $dm= \nu_{over} r^2 d\Omega=\frac{2}{\gamma-1} \nu r^2 d\Omega$. Substituting here the
equation for $d\Omega$, we obtain after simple calculations:
\begin{equation}
 \label{d3}
 w_2 = \dfrac{2\pi}{\gamma-1} G\nu R_{vir} \dfrac{a^2}{\upsilon^2_\infty}
\end{equation}
It is interesting to compare the contributions $\mathfrak F_1$ and $\mathfrak F_2$. Dividing
(\ref{b3}) by (\ref{d3}), we obtain:
\begin{equation}
 \label{d4}
 \dfrac{w_1}{w_2} = \dfrac{(\gamma-1)}{6} \dfrac{A(c_{vir})\ln (c_{vir}/2)}{c_{vir}} \dfrac{|\phi_0|}{a^2}
\end{equation}
One can see that the ratio ${w_1}/{w_2}$ grows with $M_{vir}$. The equality is reached at
\begin{equation}
 \label{e1}
 M_{vir} = \dfrac{12 a^3\sqrt{2\pi\Delta_{vir}\rho_u}}{\left(G (\gamma-1) \ln (c_{vir}/2)\right)^{3/2}}
\end{equation}
 For instance, if
$c_{vir}\equiv 70$,  $a=1 \text{km}/\text{s}$, $\gamma=4/3$, it corresponds to $M_{vir}\simeq
3\cdot 10^{6} M_\odot$. So component ${\mathfrak F_2}$ dominates for all the clumps except for the
hugest ones.

It is easy to see that in the opposite case, when $\upsilon_\infty \ll a$, the flow pattern is
determined by the ratio between the minimum of the clump potential and $a^2$. If $|\phi_0|<a^2$,
the drag force acting on the clump is negligible. In fact, in this case the flow is everywhere
subsonic. Then, because of D'Alembert's paradox, the drag force is totaly created by the viscosity
of the stream substance, that is always very small in astrophysical systems. $|\phi_0|>a^2$ for
$M_{vir}> 2.5\cdot 10^{3} M_\odot$, if we adopt $c_{vir}\equiv 70$,  $a=1 \text{km}/\text{s}$.
Consequently, small clumps do not interact with the stream.

If $|\phi_0|$ is big enough, the gas flow becomes supersonic at some region, and a shock wave may
appear there. The wave occurrence leads to a strong enhancement of the drag force. By analogy with
the previous case we can conclude that only the stream lines crossing the shock wave give a
significant contribution into the drag force. The exact hydrodynamical solution of the task can
hardly be found. However, we can easily estimate the force. It is reasonable to assume that the
shock may appear only on the lines passing through the region where $|\phi|>a^2$. Its radius $r_c$
is determined by equation (\ref{c1})
\begin{equation}
 \label{d5}
 \phi(r_c) = -\dfrac{G M_{vir}}{r_s A(c_{vir})} \dfrac{\ln(r_c/r_s+1)}{r_c/r_s}=a^2
\end{equation}
A good approximation for $r_c$ is
\begin{equation}
 \label{d6}
 r_c = r_s \dfrac{\phi_0}{a^2} \ln\left(\dfrac{\phi_0}{a^2}\right)
\end{equation}
Now we may roughly estimate the force acting on the clump as $\dfrac{\nu \upsilon^2_\infty}{2}\cdot
\pi r^2_c$
\begin{equation}
 \label{d7}
 \mathfrak F = \dfrac{\pi}{2} \nu \upsilon^2_\infty r^2_s \left(\dfrac{\phi_0}{a^2}\right)^2
 \ln^2 \left(\dfrac{\phi_0}{a^2}\right)
\end{equation}
As we can see, the drag force increases as $\upsilon^2_\infty$ if $\upsilon_\infty \ll a$ and
decreases as $\upsilon^{-2}_\infty$ if $\upsilon_\infty \gg a$, Consequently, we may assume that
the force mounts to the maximum value when $\upsilon_\infty \simeq a$.

\begin{figure}
 \resizebox{1.1\hsize}{!}{\includegraphics[angle=270]{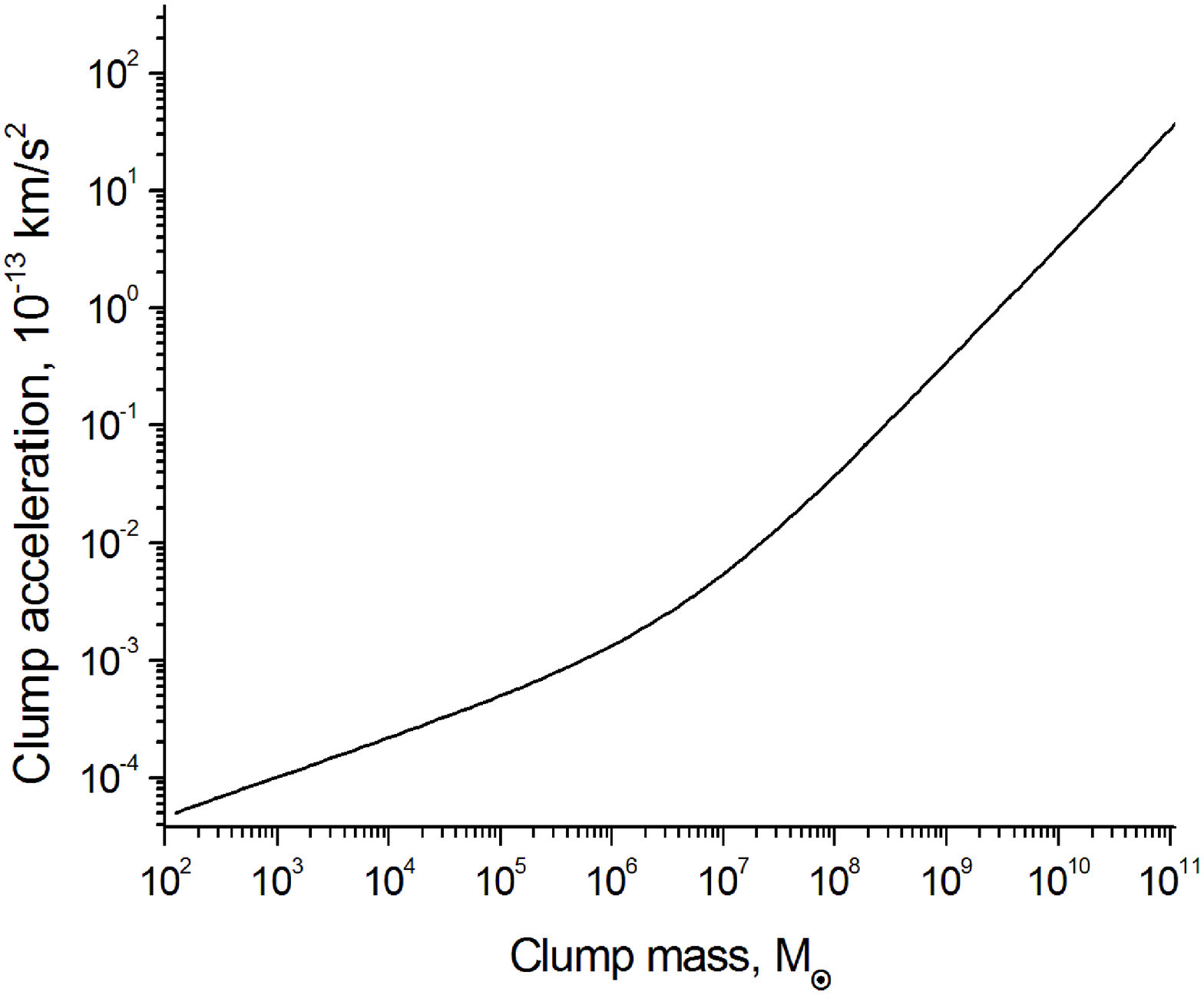}}
 \caption{The clump acceleration dependence on the clump mass. The force acting on the clump is determined by the sum
 of (\ref{b3}) and (\ref{d3}). We take $\nu=10^{-25}\text{g}/\text{cm}^3$, $\upsilon_\infty=1\; \text{km}/\text{s}$,
 $a=10\; \text{km}/\text{s}$, $\gamma=4/3$, $c_{vir}=70$.}
 \label{fig1}
\end{figure}

Reviewing equations (\ref{b3}), (\ref{d3}) and (\ref{d7}), we can make an another very common
conclusion: the drag effect always grows with the clump mass. In fact, it is easy to see that the
clump acceleration $w=\mathfrak F/M_{vir}$ increases with $M_{vir}$ in each case considered.
However, the shape of the growth is different. In the most important instance of a supersonic
stream and large clumps the acceleration increase is relatively slow ($w\sim M^{1/3}_{vir}$) if the
clump mass is below the limit (\ref{e1}). For heavier clumps the acceleration is proportional to
the mass ($w\sim M_{vir}$), see Fig.~\ref{fig1}.

\section{Discussion}
Now we can apply the obtained results to real astrophysical systems. As the first illustration, let
us consider a product of collision between two galaxy clusters, known as "bullet cluster" 1E
0657-56. During the collision hot x-ray emitting gas penetrated through the dark matter halo.  We
adopt the collision velocity (i.e. $\upsilon_\infty$) to be equal to $4700\; \text{km}/\text{s}$,
$\nu=1.5\cdot 10^{-27} \text{g}/\text{cm}^3$ ($n\simeq 10^{-3}\; \text{cm}^{-3}$), $a\simeq 1000\;
\text{km}/\text{s}$ \citep{bullet}. It easy to see that the limit (\ref{e1}) in this case cannot be
reached at any reasonable $M_{vir}$, and we should use equation (\ref{d3}). For a massive clump
($M_{vir}= 10^{8} M_\odot$) we obtain $w \simeq 5\cdot 10^{-17}\;\text{m}/\text{s}^2$. The clump
velocity increment in the characteristic time of the collision $\sim 10^8$~years is
$\delta\upsilon\sim 14\; \text{m}/\text{s}$, which is completely negligible. Consequently, the
mechanism considered cannot disturb the dark matter structure during galaxy or galaxy cluster
collisions because of high relative speeds of the objects.

The mechanism also does not result in a significant momentum exchange between the interstellar
medium in the disk of our Galaxy and the dark matter. In fact, let us consider a huge clump of mass
$10^6 M_\odot$ (heavier clumps hardly can be present in the disk of the Galaxy now). We can use
equation (\ref{d3}). Hot interstellar clouds in the disk have the temperature $\sim 10^4$~K and the
density $\nu\simeq 2\cdot 10^{-28}\text{g}/\text{cm}^3$, cold ones --- the temperature $\sim 100$~K
and the density $\nu\simeq 2\cdot 10^{-28}\text{g}/\text{cm}^3$ \citep{interstellar}. However, the
drag force acting on the clump does not depend on the gas temperature: indeed, the interstellar gas
is in hydrodynamical equilibrium, and its density $\nu\propto T^{-1}\propto a^{-2}$. Consequently,
the multiplier $\nu a^{2}$ in (\ref{d3}) remains constant. For definiteness sake we consider a hot
cloud ($a\simeq 10\; \text{km}/\text{s}$). Taking the velocity difference between the halo and the
disk rotations to be equal to $\upsilon_\infty\simeq 150\; \text{km}/\text{s}$, we obtain $w \simeq
8\cdot 10^{-15}\;\text{m}/\text{s}^2$. The age of the Galaxy is $\sim 10^{10}$~years and even if
the clump has never left the disk, its velocity change does not exceed $\delta\upsilon=30\;
\text{m}/\text{s}$.

Let us consider formation of our Galaxy \citep{galaxy}. As of now, the details of this process are
unclear.  The Galaxy was likely formed as a result of the protogalaxy collapse, which size was at
least an order larger than the present radius of the Galaxy disk. A merging of smaller structures
played an important role in the process.

The collapse initially comes about almost as a free-falling. Subsequently, however, dissipation
processes in the gas resulted in the separation of the normal and the dark components. The
collisionless dark matter stopped collapsing and formed an extensive and almost spherically
symmetric halo. The gas component lost its energy via emission, kept on compressing and formed a
compact thin disk. Consider if a significant momentum transition could take place during the
collapse.

We set the characteristic radius where the dark matter detached to be equal to $\sim 30\;
\text{kpc}\simeq 10^{23}\;\text{cm}$, the baryon mass of the Galaxy --- to
$3\cdot10^{11}\;M_\odot\simeq 6\cdot 10^{44}\;\text{g}$. Then the average density of the gas was
equal to $\nu\simeq 1.5\cdot 10^{-25}\;\text{g}/\text{cm}^3$. We adopt the protogalaxy temperature
$\sim 10^4$~K, which corresponds to sound speed $a\simeq 10\;\text{km}/\text{s}$. Substituting this
value (and $\gamma=4/3$, $c_{vir}=70$) to (\ref{e1}), we obtain $M_{vir}= 3\cdot 10^{9} M_\odot$.
Below this mass we should use equation (\ref{d3}). We adopt the time of separation of the baryon
from the dark matter to be equal to $t=0.1 t_{col}$, where $t_{col}$ is the characteristic time
scale of the collapse of the Galaxy ($t_{col}\sim 10^9$~{years}). Substituting all the values into
(\ref{d3}), we finally obtain:
\begin{multline}
 \label{f1}
 \delta\upsilon = \dfrac{2\pi}{\gamma-1} G\nu  R_{vir}
 \dfrac{a^2}{\upsilon^2_\infty}\simeq\ldots\\
 \simeq 0.5 \left(\dfrac{a}{\upsilon_\infty}\right)^2 \left(\dfrac{M_{vir}}{M_\odot}\right)^{\frac13}
 \left[\dfrac{\text{km}}{\text{s}}\right]
\end{multline}
Since the velocities of the dark and baryon matters were identical at the beginning of the
collapse, and the drag force reaches its maximum when the velocity difference is of the order of
the sound speed, we can estimate the minimal mass of the clump that could be dragged by the baryon
matter by substituting $\upsilon_\infty=a$ into (\ref{f1}).
\begin{equation}
 \label{f2}
 \delta\upsilon_{max}\simeq 0.33 \left(\dfrac{\nu}{10^{-25}\;\text{g}/\text{cm}^3}\right)
\left(\dfrac{M_{vir}}{M_\odot}\right)^{\frac13}
 \left[\dfrac{\text{km}}{\text{s}}\right]
\end{equation}
Taking $\delta\upsilon=50$~{km/s} as a significant velocity change, we obtain $M_{vir}\sim 10^6
M_\odot$. A velocity change $\delta\upsilon=10$~{km/s} corresponds to $M_{vir}\sim 10^4 M_\odot$.
In all the cases the clump should me massive enough to be carried along by the gas. On the other
hand, even much heavier clumps should be present during the Galaxy formation ($M_{vir}\ge 10^9
M_\odot$).

\begin{figure}
 \resizebox{1.1\hsize}{!}{\includegraphics[angle=270]{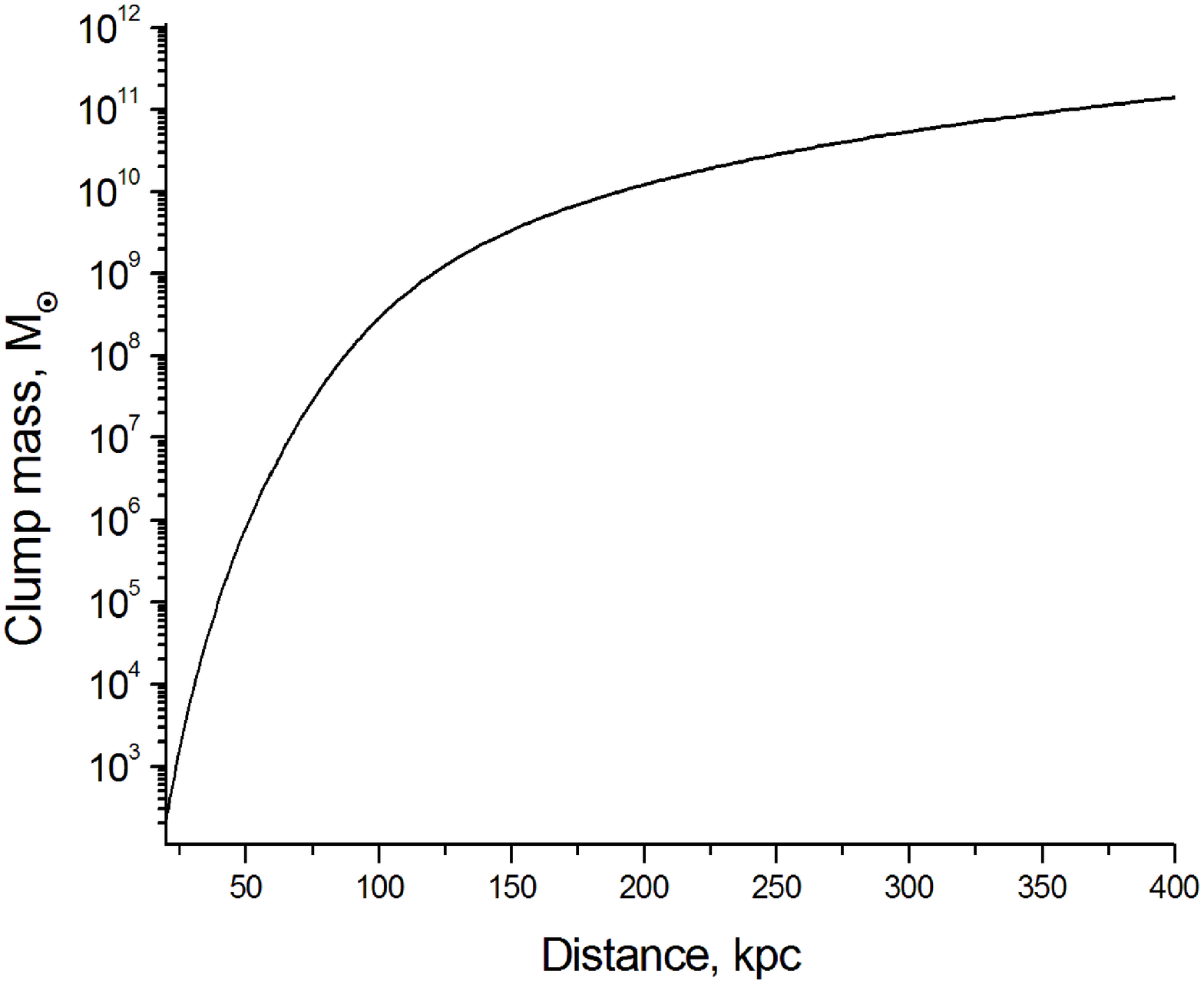}}
 \caption{Minimal mass of a clump, which velocity could be changed by the drag force on more than $\delta\upsilon=10\;
 \text{km}/\text{s}$ during the evolution of the Galaxy, depending on the distance from the
 galactic centre. The force acting on the clump is determined by the sum
 of (\ref{b3}) and (\ref{d3}).}
 \label{fig2}
\end{figure}

The momentum transmission from the gas to the dark matter leads to at least three important
consequences. First, the large clumps, carried along by the gas, were destroyed by tidal forces.
However, the dark matter of the mergers flowed into the halo, forming a more compact, oblate, and
faster rotating substructure (usually called the thick disk). This effect has been already
discovered numerically by \citep{thickdisk}.

Second, the thick disk should be more clumpy than the halo. In fact, the gas collapse could carry
along only the clumpy component of the halo, while the homogeneous remained almost spherically
symmetric. Of course, the gas could drag only large clumps that were later destroyed by tidal
forces. However, the large mergers contained a hierarchical system of smaller clumps, which were
moved in such a way to the thick disk. These low-massive clumps are too small to be destroyed by
tidal forces, and the thick disk turns out to be enriched with them. Meanwhile, these are the small
clumps that give the main contribution to the possible dark matter annihilation signal, even if
they make up only a small fraction of the dark matter. Thus we can expect that if some indirect
dark matter search find any signal, the so-called boost factor $C\equiv
\langle\rho^2_{dm}\rangle/\langle\rho_{dm}\rangle^2$ will be higher in the thick disk.

Third, only the clumps situated relatively close to the Galaxy centre could be carried along with
the gas. Indeed, the average density of the protogalaxy increased with its collapse as $r^{-3}$.
Consequently, the drag force also rapidly decreases with the radius. By analogy with equation
(\ref{f2}) we can estimate the minimal mass of a clump, which velocity could be changed by the drag
force on more than $\delta\upsilon=10\; \text{km}/\text{s}$ during the protogalaxy collapse. We
determine the force acting on the clump as the sum of (\ref{b3}) and (\ref{d3}), substituting there
$\upsilon_\infty=a\simeq 10\; \text{km}/\text{s}$ and the above-mentioned parameters of the
protogalaxy. The results are represented in Fig.~\ref{fig2}. We can see that above $r= 150\;
\text{kpc}$ the velocity disturbation is less than $10\; \text{km}/\text{s}$ even for the hugest
clumps $M_{vir}\ge 2\cdot 10^9 M_\odot$. It means, that the dark matter structure at the outskirts
of the halo remained untouched, and the big clumps from there were not transported to the thick
disk. Consequently, the boost factor of the outer regions should be higher.

It is interesting to compare the described above effect with the efficiency of direct collisions of baryons with the
dark matter particles. The number of collisions occurring in a unit volume in a unit time is $\sigma n_b n_{dm}
\upsilon_\infty$, each of them transmits a momentum $\sim m_b \upsilon_\infty$ on the average. Here we have symbolized
the concentration and mass of the dark matter particles and baryons by $n_{dm}, m_{dm}$ and $n_b, m_b$ respectively,
$\sigma$ is the cross-section of a dark matter particle scattering on a baryon. The mass of dark matter in the volume
considered is $m_{dm} n_{dm}$, and its acceleration is:
\begin{equation}
\label{a29}
 w\sim \dfrac{\sigma n_b n_{dm} \upsilon_\infty\cdot m_b \upsilon_\infty}{m_{dm} n_{dm}}=\sigma n_b \upsilon^2_\infty
 \dfrac{m_b}{m_{dm}}
\end{equation}
For a neutralino of mass $100$~{GeV} a cross-section higher than $10^{-44}\;\text{cm}^2$ is now
experimentally excluded \citep{xenon}. The effect grows with the increasing of the relative speed
of the colliding objects. However, even for the above-mentioned example of cluster collision, where
the relative speed is the largest, acceleration (\ref{a29}) does not exceed $2\cdot
10^{-34}\;\text{m}/\text{s}^2$, which is at least eight orders lower than the acceleration produced
by the mechanism under consideration.

\section{Acknowledgements} This work was supported by the RFBR (Russian Foundation for Basic
Research, Grant 08-02-00856).

\label{lastpage}

\begin{thebibliography}{99}

\bibitem[\protect\citeauthoryear{Ahn \& Komatsu}{2005}]{ahn2005} Ahn, K., and Komatsu, E., {\it
Phys. Rev. D}, {\bf 71}, 021303(R), (2005).

\bibitem[\protect\citeauthoryear{Aprile et al.}{2011}]{xenon} Aprile, E., Arisaka, K.,
Arneodo, F., et al., {\it Phys. Rev. Letters}, {\bf 107}, 131302, (2011).

\bibitem[\protect\citeauthoryear{Berezinsky, Dokuchaev, \& Eroshenko}{2006}]{berezinsky2006}
Berezinsky, V., Dokuchaev V., and Eroshenko, Yu., {\it Phys. Rev. D}, {\bf 73}, 063504, (2006).

\bibitem[\protect\citeauthoryear{Bertone, Hooper, \& Silk}{2005}]{bertone2005} Bertone, G.,
Hooper, D., Silk, J., {\it Physics Reports}, {\bf 405}, Issue 5-6, 279, (2005),
arXiv:hep-ph/0404175

\bibitem[\protect\citeauthoryear{Binney \& Merrifield}{1998}]{galaxy} Binney, J., Merrifield, M.,
{\it Galactic Astronomy}, Princeton University Press, (1998)

\bibitem[\protect\citeauthoryear{Binney \& Tremaine}{2008}]{binneytremaine}
Binney, J., \& Tremaine, S., {\it Galactic Dynamics: Second Edition},  Princeton University Press,
Princeton, NJ USA, (2008).

\bibitem[\protect\citeauthoryear{Bondi}{1952}]{Bondi52} Bondi, H., {\it MNRAS},  {\bf 112}, 195,
(1952).

\bibitem[\protect\citeauthoryear{Bondi\& Hoyle}{1944}]{BondiHoyle} Bondi, H., Hoyle, F., {\it
MNRAS},  {\bf 104}, 273, (1944).

\bibitem[\protect\citeauthoryear{S{\'a}nchez-Salcedo\& Brandenburg}{1999}]{brandenburg}
S{\'a}nchez-Salcedo, F.~J., \& Brandenburg, A.\ 1999, {\it Ap.J. Letters}, {\bf 522}, L35, (1999).

\bibitem[\protect\citeauthoryear{Bullock et al.}{2001}]{bullock} Bullock, J. S., Kolatt, T. S., Y.
Sigad, Y., Somerville, R. S., Kravtsov, A. V., Klypin, A. A., Primack, J. R., Dekel, A., {\it
MNRAS} {\bf 321}, 559, (2001).

\bibitem[\protect\citeauthoryear{Chandrasekhar}{1943}]{chan} Chandrasekhar S., {\it Ap.J.}, {\bf
97}, 255, (1943).

\bibitem[\protect\citeauthoryear{Gorbunov \& Rubakov}{2008}]{gorbrub1} Gorbunov, D. S., and
Rubakov, V. A.,   {\it Introduction to the Early Universe theory. The hot big bang theory.}, LKI
publishing house, Moscow,  (2008), (in russian).

\bibitem[\protect\citeauthoryear{Gorbunov \& Rubakov}{2009}]{gorbrub2} Gorbunov, D. S., and
Rubakov, V. A., {\it Introduction to the Early Universe theory. Volume 2: Cosmological
perturbations. Inflation theory.}, LKI publishing house, Moscow, (2010), (in russian).

\bibitem[\protect\citeauthoryear{Hofmann, Schwarz, \& St\"ocker}{2001}]{10-6} Hofmann, S.,
Schwarz, D. J. and St\"ocker, H., {\it Phys. Rev. D}, {\bf 64}, 083507, (2001).

\bibitem[\protect\citeauthoryear{Lipunov}{1992}]{lipunov}
 Lipunov, V.M., (1992), {\it Astrophysics of neutron stars}, Springer Verlag, New York, Berlin, 300 ð.

\bibitem[\protect\citeauthoryear{Markevitch \& Vikhlinin}{2007}]{bullet}
 Markevitch, M., Vikhlinin, A., {\it Physics Reports}, {\bf 443}, Issue 1, p. 1-53, (2007).

\bibitem[\protect\citeauthoryear{Ostriker}{1999}]{ostriker}
Ostriker, E.~C., {\it Ap.J.}, {\bf 513}, 252, (1999).

\bibitem[\protect\citeauthoryear{Read et al.}{2008}]{thickdisk} Read, J. I., Lake, G., Agertz O.,
and Debattista, V. P., {\it MNRAS}, {\bf 389}, 1041, (2008).

\bibitem[\protect\citeauthoryear{Spergel et al.}{2003}]{wmap} Spergel D. N., Verde L., Peiris H.
V. et al., {\it ApJS}, {\bf 148}, 175, (2003).

\bibitem[\protect\citeauthoryear{Spitzer}{1978}]{interstellar} Spitzer, L. , {\it Physical
Processes in the Interstellar Medium}, Wiley, (1978).

\bibitem[\protect\citeauthoryear{Zybin, Vysotsky, \& Gurevich}{1999}]{10-12} Zybin, K. P.,
Vysotsky, M. I., and Gurevich, A. V., {\it Phys. Lett. A}, {\bf 260}, 262, (1999).

\end{thebibliography}
\end{document}